# Relativistic Strophotron Free Electron Laser (RS FEL)

**Koryun Oganesyan [1,2,*], Ashot Gevorgyan [3] , Karen Ivanyan[4] and Peter Kopcansky [2]**

[1] Institute of Experimental Physics SAS, Kosice, Slovakia;
[2] A.Alikhanyan National Science Lab, Yerevan Physics Institute, Yerevan, Armenia;
[3] Institute of High Technologies and Advanced Materials, Far Eastern Federal University, Vladivostok, Russia;
[4] Lomonosov Moscow University, Moscow, Russia
* Correspondence: bsk@yerphi.am ;

**Abstract:** This study investigates possibility of realizing of Relativistic Strophotron Free Electron Laser (RS FEL) with a special focus on the stability of electron motion in such system. The FEL problem as a hard problem presents the challenge of creating electromagnetic tunable coherent radiation in wide wavelength range, especially in the THz and X-ray ranges. The paper deals with a realization of relativistic strophotron FEL. The optimal configuration of magnets has been identified, which achieved stable movement of electrons in relativistic strophotron. The results are applied to periodicity of beam motion in both transverse directions to achieve the conditions for RS FEL realization. The research findings suggest that using a special configuration of magnets can contribute to the optimization of the relativistic strophotron making movement of an electron beam , which is stable and useful for the creation of RS FEL. We highlight the creation of RS FEL, compare it with usual undulator (see below) and point out the advantage of an RS FEL under certain conditions. A scheme with quadrupole lenses is presented for the realization of a relativistic strophotron type Free electron laser. The presented scheme provides electrons stability in the transversal direction. The equations of motion are solved and the trajectories are found. It is shown, that the movement of electrons in the proposed scheme is stable in both transverse directions.

**Data access statement**: Research data supporting this publication are available from the 1612.05067v1 at located www.arxiv.org .

**Ethics statement** : All procedures performed in studies involving human participants were in accordance with the ethical standards in the institutional research committee.

## 1. Introduction

Free electron lasers are powerful tools for scientific research, which are coherent, powerful, and tunable sources of radiation. Currently, they are widely used in medicine, biophysics, plasma heating, materials science, etc.

In free electron lasers, the main element is a magnetic undulator, a magnetic field of which periodically depends on the longitudinal coordinate [1-9]. FELs are rapidly developing field of science (see e.g. [10-31]).

The first free-electron laser was developed by Madey et coworkers [1-4] . They used Motz technology based on magnetic wigglers [5,6]. FELs allow scientists to move into the X- ray and THz frequency ranges, where there are no other sources of coherent radiation.

Now different X ray FELs exist: Linear Coherent Light Source (LCLS) at SLAC National Accelerator Laboratory [12-15], Free electron Laser in Hamburg (FLASH) [16], European X-ray free electron laser (EuXFEL) [17], SPring-8 Compact SASE Source (SCSS) in Japan [18], SwissFEL at the Paul Scherrer Institute in Switzerland [19], PAL-XFEL (Pohang Accelerator Laboratory X-ray Free-Electron Laser) in Korea [20].

Other systems also exist allowing transverse oscillations of electrons, and consequently, there are other possibilities for creating FELs (Compton FEL, Cherenkov FEL, Smith-Purcell FEL, FEL on transition radiation, FEL without inversion [21-24] et al.

One such system is the strophotron [25-29]. A strophotron is a system with a parabolic potential (electric or magnetic) in the transverse direction of the system, into which electrons pass in the longitudinal direction and oscillate in the transverse direction.

The resonance frequency of the strophotron depends on the initial transversal electron coordinate $x_0$, small angle of electron injected in the strophotron, to the longitudinal direction $\alpha << 1$, relativistic factor $\gamma$, and frequency of transverse oscillations in electrostatic and in magnetic strophotrons $\Omega$.

The strong dependence of resonance frequency exists only in the strophotron and does not exist in the undulator, and in the relativistic strophotron spectral distribution of spontaneous radiation and gain depends on the parameter $\left(\alpha^2 + x_0^2\Omega^2\right)$ [25]. This means that they should be averaged they over the initial coordinate If it is possible to create conditions such as $\left(\alpha^2 + x_0^2\Omega^2\right) = const$, there will be no averaging and the gain of the strophotron will be an order of magnitude higher than that of the undulator. There is no such dependence in undulator. The resonance frequency does not depend on the initial coordinate of the electron [7].

A single quadrupole lens does not meet the requirements for transverse stability; in one direction the electrons oscillate, and in the other the beam decays. However, for the RS FEL to operate, beam correction in both transverse directions is necessary. We need to find systems with magnetic configurations, that provide stability of electron beam in the a strophotron.

In the present article self - consistent scheme is described for the correction of electron trajectories using additional short quadrupole lenses providing electron beam stability and oscillation periodicity in both transverse coordinates.

## 2 Methods

### Quadrupole Lenses Scheme

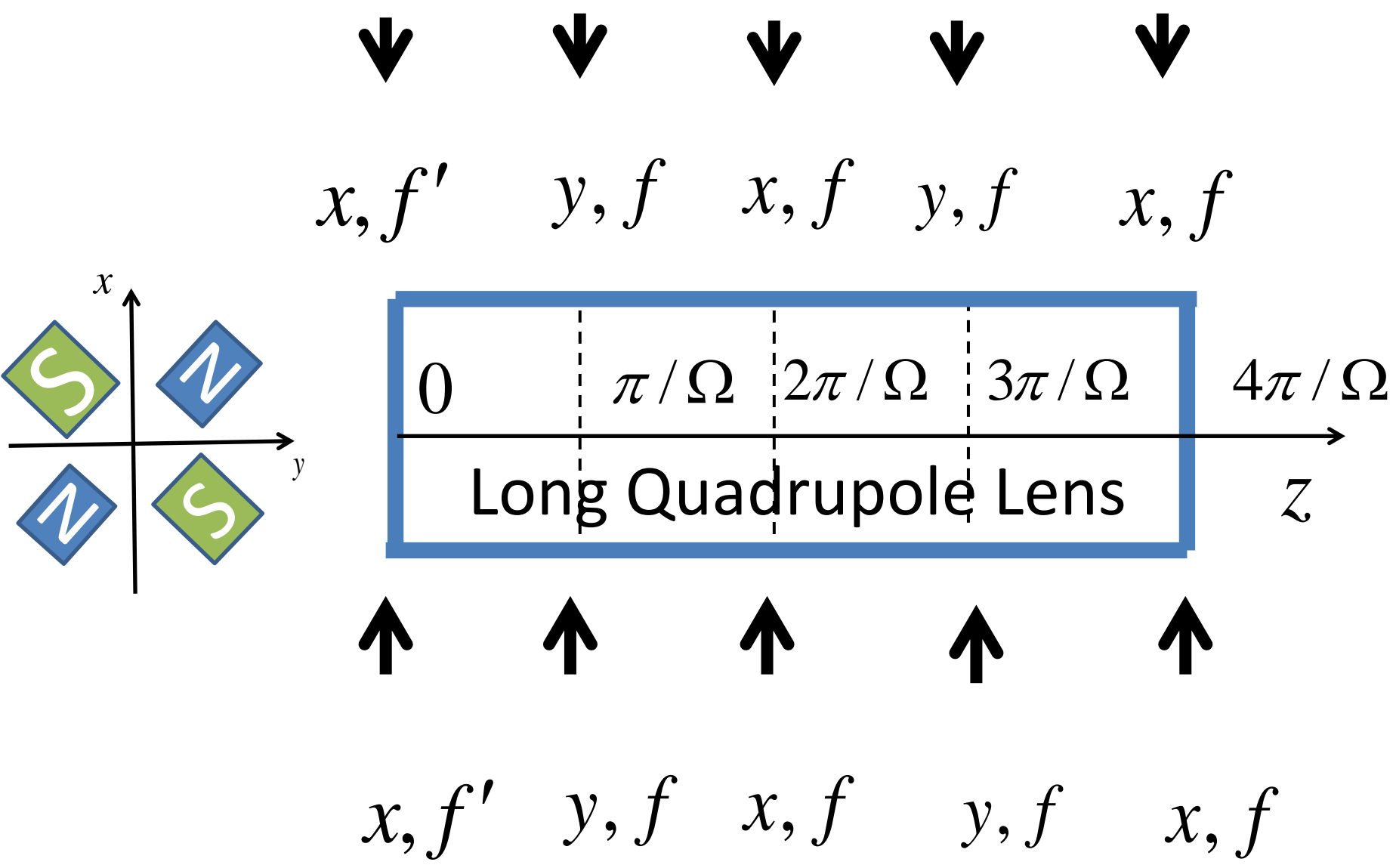

Fig. 1. Basic and corrective lenses layout. Arrows in the figure indicate the positions of corrective lenses along the $Oz$ axis of the system at a distance $\pi / \Omega$ from each other. Quadrupole magnets cross section perpendicular to the system axis Oz is presented on the left side of Figure.

$\Omega = \frac{1}{d}\sqrt{2eA_0 / \varepsilon_z}$ is the frequency of the electron oscillation in the $x$ directions of the basic quadrupole lens. Indexes $x$ and $y$ show the transverse directions in which the corresponding lens focuses. The corrected lenses are assumed to be short, so their length $l$ is much less than the focal length $f$. $f' = 2f$ is the focal length of the first lens located at the entrance ( $z = 0$ ), and $f_0$ is the focal length of the basic quadrupole lens $f_0 = \pi / 2\Omega << L$.

Electron motion in the magnetic lens with vector potential $\mathbf{A} \parallel Ox$, $A_z = -\frac{A_0}{d^2}\left(x^2 - y^2\right)$ the solutions of the equations of electron motion are

$$x'' + \Omega^2 x\left[v_0 \cos\alpha + \frac{1}{2}\Omega^2\left(x^2 - x_0^2\right) - \frac{1}{2}\Omega^2\left(y^2 - y_0^2\right)\right] = 0 \qquad (5)$$

$$y'' - \Omega^2 y\left[v_0 \cos\alpha + \frac{1}{2}\Omega^2\left(x^2 - x_0^2\right) - \frac{1}{2}\Omega^2\left(y^2 - y_0^2\right)\right] = 0. \qquad (6)$$

In the approximation of small anharmonicity $|\alpha| << 1$, $|x_0|\Omega << 1$, and $|y_0|\Omega << 1$ can be neglected. Solutions of equations (5) and (6) at $|x'| << 1$, and $|y'| << 1$ are

$$x(t) = x_0 \cos\Omega t + \frac{x_0'}{\Omega}\sin\Omega t, \qquad y(t) = y_0 \,\mathrm{ch}\,\Omega t + \frac{y_0'}{\Omega}\mathrm{sh}\,\Omega t,$$

where $x_0$ and $x_0' \approx \alpha$ are the initial coordinate and velocity along the $Ox$ axis, respectively, and $\alpha \approx \sin\alpha = p_\perp / p$ is the electron entry angle in the strophotron.

where now $y_0$ and $y_0'$ are initial coordinate and velocity along $y$ axis.

$\Omega = \frac{1}{d}\sqrt{2eA_0/\varepsilon_z}$, where $A_0$ and $2d$ are the height and width of the potential through in direction $Ox$.

Those equations show that there is a focusing in $Ox$ direction and there are possible electron oscillations, and there is defocusing in direction $Oy$ and occurs beam decay of the beam. This means that for beam stabilization its correction is needed.

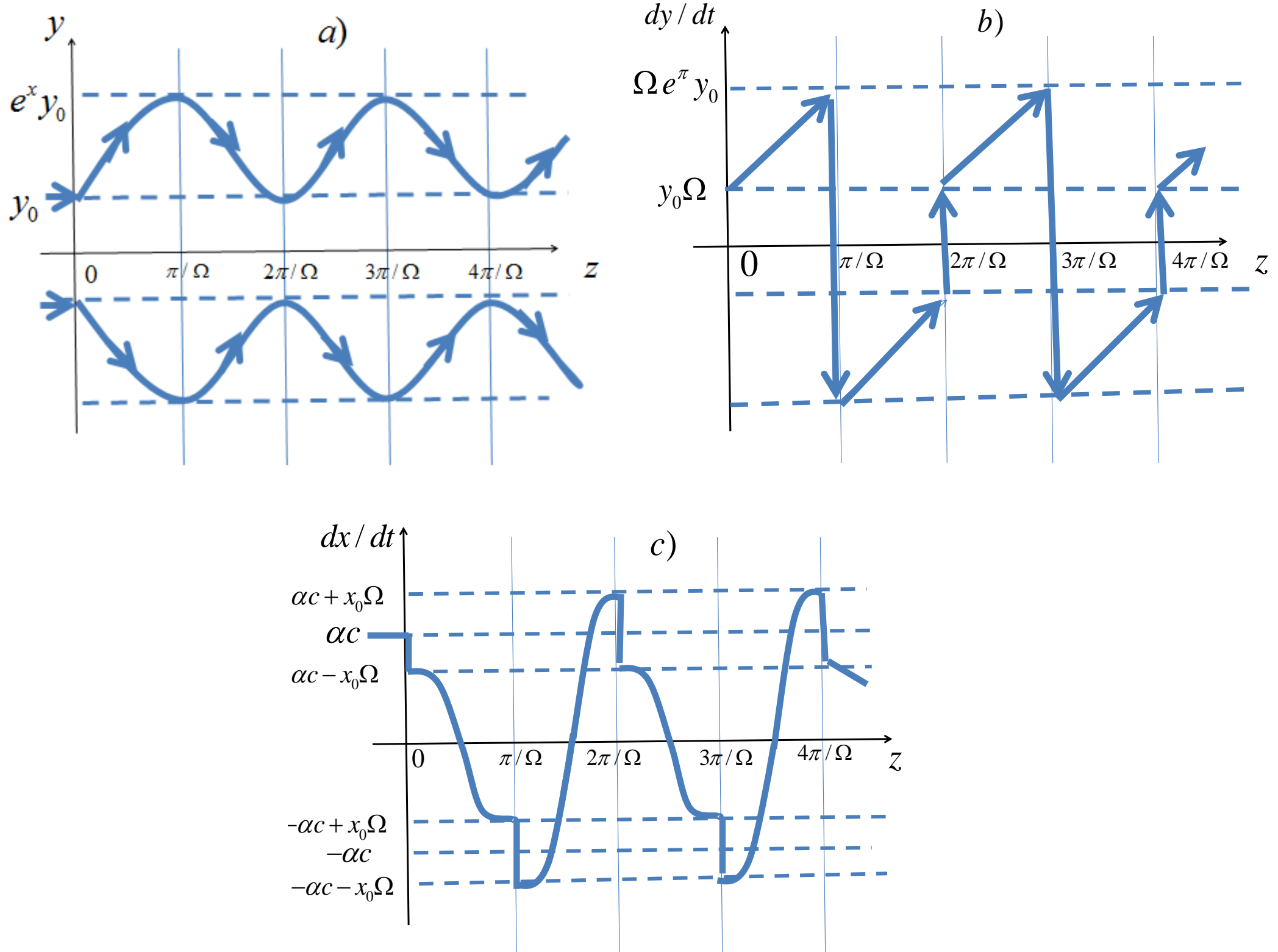

Fig.2. Electron trajectories in relativistic strophotron. Dependences of transverse coordinate $y$ (a), velocities $y'$ (b) and $x'$ (c) on longitudinal coordinate $z \approx ct$.

Well known hard focusing with crossed quadrupole lenses is used for beam stabilization. The arrows in the figure indicate the positions of corrective lenses along the $Oz$ axis of the system at a distance $\pi / \Omega$ from one another. Indexes $x$ and $y$ show the transverse directions in which the corresponding lens focuses. The corrected lenses are assumed to be short, so their length $l$ is much less than the focal length $f$. This assumption allows us to not analyze in detail the electron motion in the corrective lenses, describing their action instantaneous change in a component of the electron velocity on the $\pm x_i f$, (minus, plus) $y_i f$, where $x_i$ and $y_i$ values of the transverse coordinate at the intersection of the electron $i$-th corrected lens.

Focal lengths of all corrected lenses (except first) we accept to be

$$f = \frac{1}{2\Omega} = \frac{f'}{2} = \frac{f_0}{\pi}, \tag{7}$$

where $f' = 2f$ is the focal length of the first lens located at the entrance ($z = 0$), and $f_0$ is the focal length of the basic quadrupole lens $f_0 = \pi / 2\Omega << L$.

This choice provides electron motion periodicity on both transverse coordinates (see Fig.2). The electron velocity in the direction of $Oy$ , $y'$ changes its direction(sign) during every correction (Fig.2. a,b), the electron velocity in the direction $Ox$ , $x'$ at the intersection of corrective lenses abruptly changed to $\pm 2x_0\Omega$ , and $\pm x_0\Omega$ at the intersection of the first lens (see Fig.2).

These abrupt changes in speed $x'$ are small compared with its initial value $x' \approx \alpha$ if the condition holds

$$\Omega \Delta x_0 << \alpha \qquad (8)$$

where $\Delta x_0$ is the transverse size of the beam in the $Ox$ direction corresponding to the electrons flying in the vicinity of the $Oz$ axis providing an effective contribution to the resonant radiation and equal $\frac{4}{\Omega}\sqrt{\lambda/(\pi L)}$, where $L$ is the system length, and $\lambda$ - is the radiation wavelength.

## 3. Results

If the necessity of correction and focusing of the beam in the $Oy$ direction is evident, the need for focus adjustment in the $Ox$ direction is associated with the periodicity of the oscillations and the absence of an accumulated change of velocity $x'$. Under condition (8) the jumps in velocity $x'$ are small and do not essentially affect the nature of the oscillations $x(t)$ , the Fourier expansion of which has a the form

$$x(t) = x_0 \cos \Omega t + \frac{\alpha}{\Omega} \sin \Omega t - \frac{2x_0}{\pi}\left(1 - 2\sum_{k=1}^{\infty} \frac{1}{4k^2 - 1} \cos 2k\Omega t\right). \qquad (9)$$

Under the selected installation scheme of corrective lenses movement in a second transverse coordinate y is also periodic with the same period $2\pi/\Omega$ if $y_0' = 0$. Condition $y_0' \neq 0$ violates periodicity. During one period $2\pi/\Omega$, $y$ increases on $y_0'/\Omega$. The deviation from the strict periodic motion in $y$ means that for $y_0' \neq 0$ is a certain degree of instability. This instability is small under condition

$$\frac{L}{\lambda}\frac{y_0'}{\Omega} = \frac{L y_0'}{2\pi} << d_{ey} \qquad (10)$$

where $L$ is the system length, and $\lambda = 2\pi / \Omega$. Condition (10) indicates that the increase in $y$ independence on $y_0' \neq 0$ on the entire length $L$ should be less than the transverse dimension of the beam in the y direction $d_{ey}$.

With values $L = 300cm$, $d_{ey} = 1cm$ the condition (10) gives

$$y_0' << \frac{2\pi d_{ey}}{L} = 2\times 10^{-2} \tag{11}$$

The condition of used approximation has a form

$$e^{\pi} d_{ey} < D_y,\ 1/\Omega, \tag{12}$$

where $D_y$ is a lens aperture in the Oy direction.

## 4. Discussion

Finally, the motion of the electrons in the system with a quadrupole lenses is stable. A small deviation in the focal distance $\delta f << f$ does not lead to instability in the movement in the $x$ direction, it leads to an increase of $y$ coordinate, so $\delta y_0 / y_0 \sim N \frac{\delta f}{f}$, where $y_0$ is coordinate $y$ at the entrance in the system, $\delta y_0$ is increased on the system length, $N$ - is the number of periods, and $f$ is the lens focal length. The condition of a small deviation along $y$ is $\delta y_0 / y_0 < 1$, which is equivalent to the requirement $\frac{\delta f}{f} < \frac{1}{N}$.

Under conditions $\left(\alpha^2 + x_0^2\Omega^2\right) = const$ a gain in relativistic strophotron type FEL is an order of magnitude higher than in a usual FEL with undulator.

## 5. Conclusions

A relativistic strophotron type FEL scheme was presented in this paper. It consists of a long quadrupole magnetic lens and several short quadrupole lenses. It is shown that the movement of electrons in both transverse directions of the relativistic strophotron is periodic and therefore can be used for the construction of relativistic strophotron type FEL.

**Funding:** “This research was funded by the EU NextGenerationEU through the Recovery and Resilience Plan for Slovakia under project No. 09103-03-V01-00052, the Slovak Academy of Sciences, in the framework of projects VEGA 2/0061/24**,** VEGA **2/0028/25** and by Slovak Research and Development Agency project No. APVV-22-0060 MAMOTEX.

**Conflicts of Interest**: “The authors declare no conflicts of interest.”